\documentclass[prl,aps,twocolumn,amsmath,amssymb]{revtex4}

\usepackage{graphicx}
\usepackage{color}

\begin{document}

\title{Superconductivity in Sr$_2$RuO$_4$-Sr$_3$Ru$_2$O$_7$ eutectic crystals}

\author{R. Fittipaldi$^{1,2}$}
\author{A. Vecchione$^{1,2}$}
\author{R. Ciancio$^{1,2,5}$}
\author{S. Pace$^{1,2}$}
\author{M. Cuoco$^{1,2}$}
\author{D. Stornaiuolo$^{3}$}
\author{D. Born$^{3}$}
\author{F. Tafuri$^{3,4}$}
\author{E. Olsson$^{5}$}
\author{S. Kittaka$^{6}$}
\author{H. Yaguchi$^{6}$}
\author{Y. Maeno$^{6}$}

\affiliation{$^{1}$ Laboratorio Regionale SuperMat, CNR-INFM
Salerno, Baronissi (Sa), Italy}

\affiliation{$^{2}$ Dipartimento di Fisica "E.R. Caianiello",
Universit\'a di Salerno, Baronissi(Sa),Italy}

\affiliation{$^{3}$ CNR-INFM Coherentia and Dipartimento di
Scienze Fisiche , Universit\'a di Napoli Federico II, Napoli,
Italy}

\affiliation{$^{4}$ Dipartimento Ingegneria dell'Informazione,
Seconda Universit\'a di Napoli, Aversa (CE), Italy}

\affiliation{$^{5}$ Department of Applied Physics, Chalmers University of Technology, University of
Chalmers, Goteborg}

\affiliation{$^{6}$ Department of Physics, Kyoto University, Kyoto
606-8502, Japan}

\begin{abstract}

Superconducting behavior has been observed in the
Sr$_2$RuO$_4$-Sr$_3$Ru$_2$O$_7$ eutectic system as grown by the
flux-feeding floating zone technique. A supercurrent flows across
a single interface between Sr$_2$RuO$_4$ and Sr$_3$Ru$_2$O$_7$
areas at distances that are far beyond those expected in a
conventional proximity scenario.  The current-voltage
characteristics within the Sr$_3$Ru$_2$O$_7$ macrodomain, as
extracted from the eutectic, exhibit signatures of
superconductivity in the bilayered ruthenate. Detailed
microstructural, morphological and compositional analyses address
issues on the concentration and the size of Sr$_2$RuO$_4$
inclusions within the Sr$_3$Ru$_2$O$_7$ matrix. We speculate on
the possibility of inhomogeneous superconductivity in the {\it
eutectic} Sr$_3$Ru$_2$O$_7$ and exotic pairing induced by the
Sr$_2$RuO$_4$ inclusions.
\end{abstract}

\maketitle


Layered strontium ruthenates are currently a subject of intense
investigation due to many fascinating properties exhibited by
Sr$_2$RuO$_4$ in the context of unconventional superconductivity
and by Sr$_3$Ru$_2$O$_7$ in relation with metamagnetic quantum
criticality\cite{Mae94,Mac03,Ike04,Gri04,Bor07}. The recent
progress\cite{Fit05} in preparing samples with eutectic
solidification of superconducting(SC) Sr$_2$RuO$_4$ and
metamagnetic Sr$_3$Ru$_2$O$_7$ have opened a novel route to
explore systems where unconventional superconductivity and
collective magnetic behavior can be matched at the nanoscale with
high interface quality. A preliminary characterization of this
type of compound, via current-voltage ($I$-$V$) measurements, has
shown an unusual temperature and field behavior of the critical
current\cite{Hoo06}. It emerged a picture of a strongly coupled
Josephson weak-link network where the Sr$_2$RuO$_4$ grains should
be responsible for the SC behavior mediated by an unusual
proximity effect through the Sr$_3$Ru$_2$O$_7$ domains. The
screening fraction has been estimated via susceptibility
measurements to be nearly 100$\%$ of the volume of
Sr$_3$Ru$_2$O$_7$ region\cite{Kit06}. The peculiar sensitivity of
the screening to small applied fields has been considered as an
evidence of an intrinsic granularity. So far, the nature of the SC
state in the eutectic turns out to be controversial, mainly for
the lack of detailed structural characterization about the
concentration and the size of Sr$_2$RuO$_4$ inclusions within
Sr$_3$Ru$_2$O$_7$ eutectic domain.

In this letter, we perform a detailed transport and structural
analysis of the Sr$_2$RuO$_4$-Sr$_3$Ru$_2$O$_7$ compound profiting
of a special sample selection within the eutectic itself. We have
investigated: i) one nominal Sr$_2$RuO$_4$-Sr$_3$Ru$_2$O$_7$
interface and ii) a macrodomain of Sr$_3$Ru$_2$O$_7$. The $I$-$V$
characteristics measured in i) shows a supercurrent $I_c$ even at
distances of $\sim$ 2000 $\mu$m far from the interface between
Sr$_2$RuO$_4$ and Sr$_3$Ru$_2$O$_7$. $I_c$ does not decay
exponentially as in a conventional proximity effect. The same
measurements performed for the case ii) give evidence of a
supercurrent flowing through the Sr$_3$Ru$_2$O$_7$ domain as well.
Systematic compositional and microstructural studies yield limits
on the distribution and size of Sr$_2$RuO$_4$ inclusions within
the Sr$_3$Ru$_2$O$_7$ matrix. Their presence cannot be more than a
few percent of the total volume and they consist of one, two or
three individual layers of Sr$_2$RuO$_4$ in a Sr$_3$Ru$_2$O$_7$
matrix interrupting the stack of the bilayered structure along the
$c$-axis. We speculate that the small amount of Sr$_2$RuO$_4$
within the Sr$_3$Ru$_2$O$_7$ cannot account for the observed
superflow at the interface and within the Sr$_3$Ru$_2$O$_7$ even
as a percolative phenomena.

The synthesis of Sr$_2$RuO$_4$-Sr$_3$Ru$_2$O$_7$ eutectic crystals
by a flux-feeding floating zone technique has been previously
reported\cite{Fit05}. Inspection with the polarized light optical
microscope (PLOM) within the $a$-$c$ planes confirmed the expected
lamellar solidification microstructure of the eutectic
Sr$_2$RuO$_4$-Sr$_3$Ru$_2$O$_7$\cite{Fit05}. The morphology of the
crystals has been investigated by scanning electron microscope
(SEM) equipped with electron back scattered (EBS) detector,
yielding no detectable precipitates. Moreover, likewise to other
single-phase ruthenates, the eutectic crystals can be cleaved in
the $a$-$b$ planes. The crystal selected for this work (hereafter
called sample 1) is basically composed both of lamellar pattern
and of large domains of the two phases. From this sample, we have
cut a piece (sample 2) composed by two macrodomains of
Sr$_2$RuO$_4$ and Sr$_3$Ru$_2$O$_7$ in contact with each other.
This is a crucial aspect for the present investigation: the
requirement is to obtain a controlled arrangement of the two
phases inside the sample. Indeed, to make sure that just one
interface is involved in the transport measurements, both the top
and bottom sides of the $a$-$c$ surface have been polished until
the same pattern of Sr$_2$RuO$_4$ and Sr$_3$Ru$_2$O$_7$ regions is
achieved. The final dimensions of the sample 2 were $5.61 \times
0.32 \times 0.66\,$ mm$^3$. The composition of the sample 2 has
been carried out both by energy dispersive spectroscopy (EDS) and
by wavelength dispersive spectroscopy (WDS). EDS analysis
performed over different areas gave evidence of the presence,
within the sample, of only two distinct phases identified as
Sr$_2$RuO$_4$ and Sr$_3$Ru$_2$O$_7$. The constant value
$n$=2\,$N$(Ru)/$N$(Sr) as obtained by WDS analysis ($N$(Ru) and
$N$(Sr) being the number of moles for Sr and Ru, respectively),
demonstrated the homogeneity of the two phases within the bulk
region. Spatially resolved WDS presented no signs of gradient
concentration of Sr$_3$Ru$_2$O$_7$ or Sr$_2$RuO$_4$ approaching
the interface. Scans of approximately 10 hours has been carried
out in the wavelengths range $0.248-2.379\,$ nm on different
regions of the sample to check the presence of spurious elements
with $Z$ ranging from 5(B) and 83(Bi). All the observed peaks have
been attributed to X-rays characteristic of Sr or Ru. No
magnetically active element has been found. The analysis on the
Sr$_2$RuO$_4$ and Sr$_3$Ru$_2$O$_7$ domains did not reveal
spurious elements within the sensitivity of the experiment.
Subsequently, from sample 2 we selected a piece of dimensions
$2.82 \times 0.32 \times 0.66\,$ mm$^3$ nominally consisting only
of Sr$_3$Ru$_2$O$_7$ phase (sample 3). This sample showed few
isolated Sr$_2$RuO$_4$ clusters on the surface of $\sim$10
$\mu$m$^2$ with an average distance among them of $\sim$30 $\mu$m.
The amount of Sr$_2$RuO$_4$ clusters visible in SEM on the surface
of sample 3 has been evaluated to be about 0.5$\%$ of the
Sr$_3$Ru$_2$O$_7$ domain.

\begin{figure}[hbt]
\includegraphics [width =0.35\textwidth]{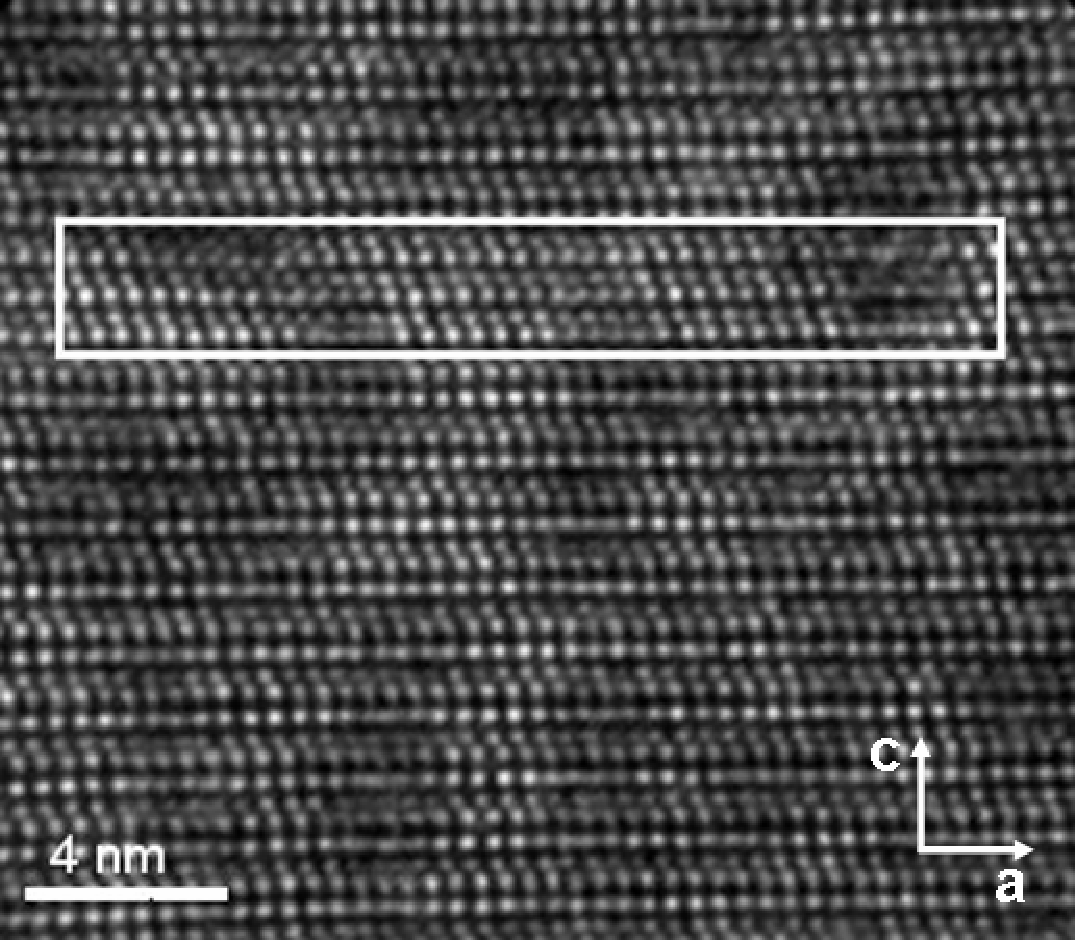}
\caption{TEM image in the $a$-$c$ plane of the Sr$_3$Ru$_2$O$_7$
domain as cut from the eutectic. The arrows indicate the area with
the atomic stacking of the Sr$_2$RuO$_4$ phase. } \label{fig:tem}
\end{figure}

To detect the presence of Sr$_2$RuO$_4$ nanodomains finely
dispersed within the Sr$_3$Ru$_2$O$_7$ layers at atomic level, we
have performed a microstructural analysis by means of transmission
electron microscope(TEM) through different areas of the sample 3.
A representative TEM image of different areas of sample 3 away
from Sr$_2$RuO$_4$ clusters is reported in Fig. \ref{fig:tem}. It
is possible to notice that the Sr$_3$Ru$_2$O$_7$ domain, as
extracted from the eutectic, is homogeneous and free of defects
over areas of several nanometers. A detailed inspection reveals
the presence of one or two unit cells corresponding to the
Sr$_2$RuO$_4$ structure, occurring as stacking faults monolayers
that, along the $c$-axis, interrupt the Sr$_3$Ru$_2$O$_7$
periodicity. The systematic scan of a large area within sample 3
indicates that such type of stacking faults defects are of
nanometric size and their amount can be estimated to be not
greater than a few percent of the total area analyzed.

As a confirmation of such dispersion of non stoichiometric defects
with respect to the Sr$_3$Ru$_2$O$_7$ structure, we traced a grid
of $20 \times 20$ points on the sample 3 to perform EDS
measurements by avoiding the few regions where the clusters of
Sr$_2$RuO$_4$ are observable. The value obtained for the Sr/Ru
ratio is $n$ = $1.304 \pm 0.012$, that turns out to be very close
to the expected one for an ideal Sr$_3$Ru$_2$O$_7$ crystal
($n$=1.333). The deviation from the ideal value in the direction
of the monolayer stoichiometry can be read as a consequence of
Sr$_2$RuO$_4$ defects.

The structure and crystalline quality of the samples were assessed
by a high resolution triple axis X-ray diffractometer, with a Cu
K$\alpha$ source equipped with a four-circle cradle. The
diffraction peaks of sample 1 were identified with the expected
$(00l)$ Bragg reflections coming from Sr$_3$Ru$_2$O$_7$ and
Sr$_2$RuO$_4$, demonstrating both the absence of any spurious
phase and the common direction of the $c$-axes of the two phases.
Pole figures proved that also the in-plane axes of the two phases
were aligned. It is worth pointing out that XRD spectrum from
sample 3 collected with a step size of 0.0002 degrees and time per
step of 150 seconds, showed the absence of any Sr$_2$RuO$_4$ peak
indicating that in the interacting volume, the Sr$_2$RuO$_4$
percentage is lower than that one detectable from the surface of
the sample.

Hence, by combining the TEM, the X-ray and the EDS/WDS analysis we
may conclude that, if present, the size of Sr$_2$RuO$_4$
inclusions dispersed in the sample 3, as one, two or three
individual layers in a Sr$_3$Ru$_2$O$_7$ matrix, is of the order
of a few nanometers in the thickness. The upper limit of the
volume fraction of these Sr$_2$RuO$_4$ layers is 5$\%$. Such
percentage can be obtained by combining the systematic TEM
analysis\cite{ciancio07} and evaluating the difference between the
measured average $n$ with respect to that expected for the ideal
Sr$_3$Ru$_2$O$_7$ stoichiometry.

\begin{figure}[hbt]
\includegraphics[width=0.45\textwidth]{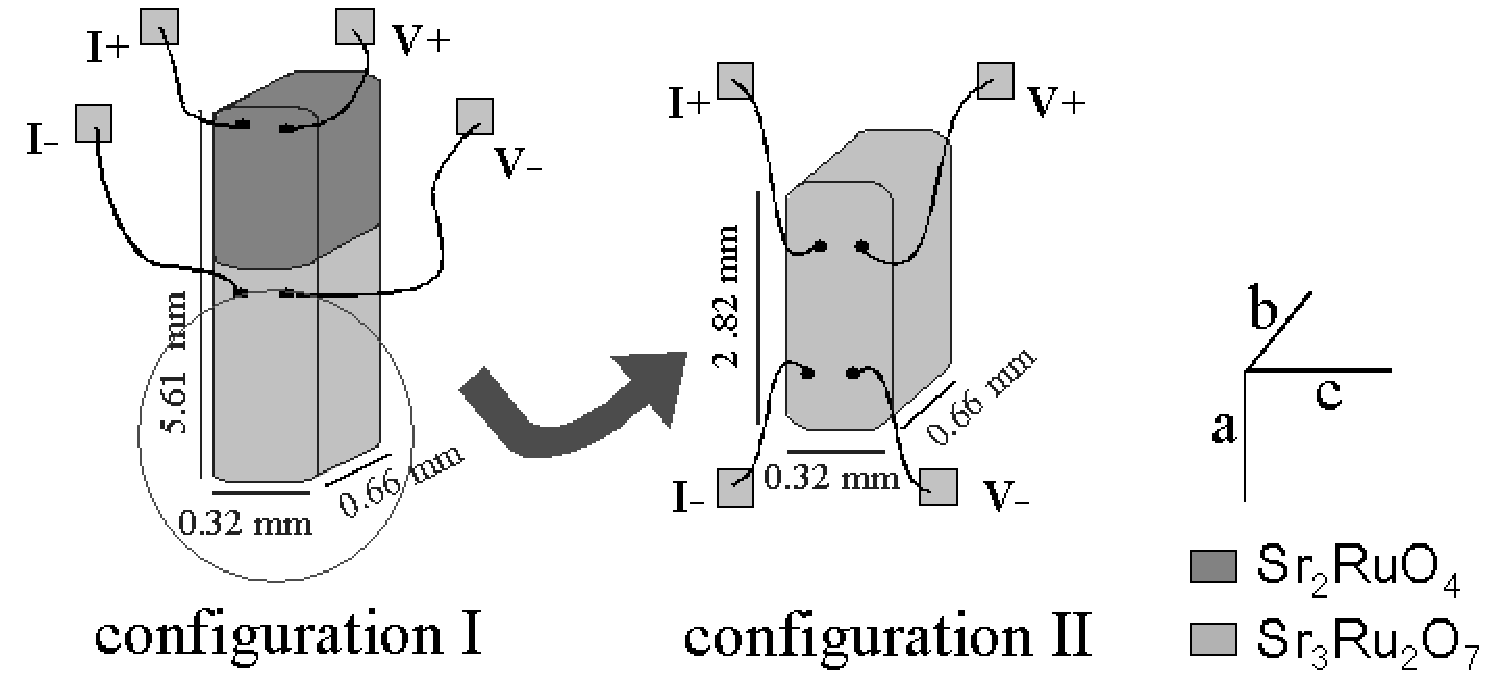}
--------------------------------------------------------------------
\includegraphics[width=0.5\textwidth]{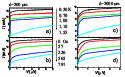}
\caption{Top: sketch of electrical leads at different positions on
sample 2 (configuration I) and on sample 3 (configuration II),
respectively. Bottom: $I$-$V$ characteristics of sample 2 measured
with contact leads in the configuration I at two different
distances from the Sr$_2$RuO$_4$-Sr$_3$Ru$_2$O$_7$ interface: a)
$H=0$ T and $T$ ranging from 1.5 K to 0.3 K, b) $T$=0.3 K and $H$
$||$ c-axis ranging from 0 to 750 Oe, $d\simeq$ 200 $\mu$m; c)
$H=0$ T and $T$ ranging from 1.5 K to 0.3 K, d) $T$=0.3 K and $H$
ranging from 0 to 750 mT, $d\simeq$ 2000 $\mu$m.}
\label{fig:IVconfig1}
\end{figure}

We have studied the $I$-$V$ characteristics of the samples 2 and 3
at low temperatures by using a $^3$He refrigerator to cool the
system down to 0.3 K. All the transport measurements have been
performed in Napoli by a standard four-point technique applying
the bias current within $a$-$c$ planes and the magnetic field
along $c$-axis. Contact leads have been placed on the crystals to
minimize uncontrollable and complicated pattern of the current
flow through the sample. The electrical leads are Al-Si wires, 50
$\mu$m in diameter, and the contact area is about 100$\times$100
$\mu$m$^{2}$. For the transport analysis we used two different
measurement configurations. The configuration I has two contacts
in the Sr$_2$RuO$_4$ region and the other two in Sr$_3$Ru$_2$O$_7$
one at a distance $d\cong $200 $\mu$m from the interface (see top
side of Fig. 2). In the same configuration, the contacts are
placed in the Sr$_3$Ru$_2$O$_7$ region at $d\cong$ 2000 $\mu$m
away from the interface. In the configuration II the contacts are
placed only on the sample 3. All the current and voltage leads
were attached using an ultrasound bonder. The measurement of the
resistance versus temperature in the configuration I below 2 K
yields a SC transition at $T_{c0}$ = 1.39 K. We do not observe
extra features as measured by a.c. susceptibility on the same
eutectic samples\cite{Kit06}. The value found is slightly lower
than the SC transition temperature in the best Sr$_2$RuO$_4$ pure
single crystals ($T_{c0}$=1.5K) and nearly the same as that one
observed in Ref.\cite{Kit06} within the Sr$_3$Ru$_2$O$_7$ region
($T_{c0}$=1.31 K) cut from the eutectic. This unexpected result is
confirmed by $I$-$V$ characteristics reported in Fig. 2a and Fig.
2b. Below $T_{c0}$ a finite well defined zero-voltage current,
$I_c$, is observed. This behaviour was reproducibly found for all
Sr$_2$RuO$_4$-Sr$_3$Ru$_2$O$_7$ junctions. The zero-voltage
current presents all the hallmarks of a typical supercurrent: it
is increased by decreasing the temperature and is reduced by
applying an external magnetic field (see Fig. 2a and Fig. 2b),
respectively. By assuming that the whole cross-sectional area was
involved in the current path, the critical current density, $J_c$,
is estimated about $7.1 $A/cm$^2$ at 0.3 K and zero applied
magnetic field. Remarkably, by placing the contact leads at a
distance $d\sim 2000 \mu$m far from the interface, a critical
current is again observed for temperature below $T_{c0}\sim 1.39$
K (see Fig. 2c). $I$-$V$ curves measured at different temperatures
and magnetic fields show a behaviour similar to the previous one
as shown in Fig. 2c and Fig. 2d. In this case, the $J_c$ value is
about $3.5 $A/cm$^2$ at 0.3 K. It proves that the supercurrent
flows on a long length scale weakly depending on how far the
contacts on the Sr$_3$Ru$_2$O$_7$ part are from the interface with
the Sr$_2$RuO$_4$ region.

In order to clarify the role played by the occurrence of an
anomalous proximity effect induced by the Sr$_2$RuO$_4$
macrodomain connected to the Sr$_3$Ru$_2$O$_7$ part, we have
performed transport measurements on sample 3 as presented in Fig.
3 (configuration II). From the behaviour of the resistance versus
temperature $R(T)$ it is possible to extract again a SC transition
at $T_{c0}$ = 1.39 K at zero applied magnetic field, showing that
a supercurrent is sustained in the Sr$_3$Ru$_2$O$_7$ macrodomain
independently on the presence of the Sr$_2$RuO$_4$ area. In Fig. 3
we present the evolution of the resistance versus temperature at
different selected magnetic field. Here, $T_c(H)$ and $H_{c2||
c}$(0) are defined as the temperature and magnetic field for which
the resistance reaches 20$\%$ of the normal state resistance. Fig.
3 shows the $H_{c2}$-$T$ phase diagram obtained from the $R(H)-T$
curves. The upper critical field at zero temperature $H_{c2}(T=0)$
can be estimated by data interpolation with a resulting amplitude
for field along the $c$-axis, $H_{c2|| c}$(0)$\simeq$ 100 mT. This
value is comparable to that of Sr$_2$RuO$_4$ single crystals with
$T_c \sim $1.4 K but substantially different from the value 1 T
obtained in the 3-K phase samples\cite{And99}.
\begin{figure}[hbt]
\includegraphics [width = 0.485\textwidth]{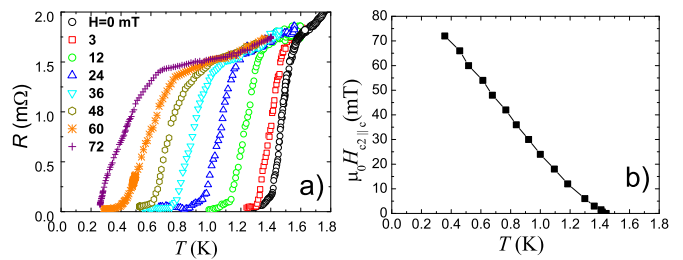}
\caption{\small {a) Resistance vs temperature of the sample 3
measured with contact leads in configuration II at different
applied field, b) extrapolated upper critical field vs
temperature.}} \label{fig:IVconfig2}
\end{figure}


One has to address the origin and the nature of the SC features
observed in the $I$-$V$ within the Sr$_3$Ru$_2$O$_7$ part of the
eutectic. Our considerations start from the basic aspects that, as
single crystal phase, the Sr$_2$RuO$_4$ is homogeneously SC while
the Sr$_3$Ru$_2$O$_7$ does not exhibit any SC feature down to very
low temperature. Further, the microstructural analysis within the
Sr$_3$Ru$_2$O$_7$, as cut from the eutectic, reveals a tiny
concentration of Sr$_2$RuO$_4$ defects either made by stacking
faults RuO monolayers, or Sr$_2$RuO$_4$ regions clusters. Hence,
the observation of a superflow, within the sample 3 and at large
distances from the interface in the sample 2, can have three
possible scenarios: i) the Sr$_3$Ru$_2$O$_7$ is not SC and the
supercurrent flows via percolation through the Sr$_2$RuO$_4$
defects as in a proximity network, ii) the Sr$_3$Ru$_2$O$_7$ has
an inhomogeneous superconductivity that supports, together with
the Sr$_2$RuO$_4$ inclusions, the supercurrent flowing within the
entire sample, iii) the Sr$_3$Ru$_2$O$_7$ is a homogeneous SC
system.

For the case i), it is difficult to account for a flow of
supercurrent in the sample 3 as a conventional percolation
phenomena through a proximity network made of Sr$_2$RuO$_4$
centers separated by normal areas of Sr$_3$Ru$_2$O$_7$. The main
problem comes from the observed distances between the nodes of the
network. Taking into account the de Haas-van Alphen measurements
\cite{Bor04} and the theoretical band structure
predictions\cite{Sin01}, it is possible to obtain the in- and out
of plane normal coherence length in the clean limit for the
Sr$_3$Ru$_2$O$_7$ of the order of $\xi_{N}^{a}\simeq
40\,T^{-1}\,$K \,nm and $\xi_{N}^{c}\simeq 2.5\,T^{-1}$\,K\,nm
($\xi_{N}^{i}=\hbar v_{F}^{i}/2\pi k_{B} T$, where $i=a,c$ are the
axis directions), respectively. Those estimations for the leaking
of Cooper pairs in the {\it normal} Sr$_3$Ru$_2$O$_7$ are not
compatible both qualitatively and quantitatively with the observed
SC features. Hence, the scenario i) can work only if some sort of
unconventional mechanism allows for an enhancing of the proximity
length within the Sr$_3$Ru$_2$O$_7$ domain.

One such possibility is related with the presence of the
nanometric Sr$_2$RuO$_4$ inclusions that can act as pairs
scattering centers for the electrons in the Sr$_3$Ru$_2$O$_7$
bands. In this case, areas of RuO monolayers, where pairing
without long-range order would occur in the $d_{xy}$ band, act
along the $c$ direction as local centers for pair exchange with
the electrons in the bands of the Sr$_3$Ru$_2$O$_7$ having
$(d_{xz},d_{yz})$ character. The latter mechanism is favored
because of the specific electronic structure and symmetry of the
$t_{2g}$ bands\cite{Zhi01}. The pairs in the RuO monolayer would
behave like effective charge impurity for the electron propagation
along the $c-$axis while the interband scattering allows for
pair-exchange between electrons in neighbor layers. It emerges a
picture similar to the charge Kondo model\cite{Tar91} and a
mechanism for long-range order between pairs in different
Sr$_2$RuO$_4$ inclusions analog of the
RKKY(Ruderman-Kittel-Kasuya-Yosida) interaction between spins. For
the charge Kondo model the effective coupling amongst two
pair/isospins, mediated by particle-particle excitations, at a
distance $d$ is $I(d)=(J^2 \rho_f/8 \pi)/d^3$ where $J$ is the
effective local pair exchange and $\rho_f$ is the density of
states at the Fermi level. Assuming a concentration $x$ of
randomly placed Sr$_2$RuO$_4$ inclusions (resonant pairing
centers) the mean-field estimation of the SC percolation
temperature gives $T_{c-eut}\cong x J^2 \rho_f$ log$(D/(x J^2
\rho_f))$ in terms of the bandwidth $D$ \cite{Mal91}. Thus, given
for the Sr$_3$Ru$_2$O$_7$ the estimated $\rho_f$($\sim$ 5
states/eV), $D$($\sim$ 2 meV)\cite{Sin01} and considering that the
rate of interlayer pair scattering $J$ is larger than $D$ ($J/D$
ranges in the interval $[4-8]$) one obtains a value of $T_{c-eut}
\sim 1 K $ for a percentage window $[2\%- 8\%]$ of Sr$_2$RuO$_4$
inclusions.

The scenario ii) and iii) invokes the possibility that part or the
entire Sr$_3$Ru$_2$O$_7$ system becomes SC. Such circumstance
requires a motivation of the lack of superconductivity in the
crystals of Sr$_3$Ru$_2$O$_7$ not grown as eutectic. While iii) is
unlikely to occur, it might be possible that the small difference
in lattice parameters between the Sr$_2$RuO$_4$ and the
Sr$_3$Ru$_2$O$_7$ results in a strain that induces pair formation
within the Sr$_3$Ru$_2$O$_7$ domain in the form of inhomogeneous
patterns. A more detailed microstructural analysis is required to
get extra insights in such issues.

In conclusion, we have shown that the Sr$_3$Ru$_2$O$_7$, as
extracted from the Sr$_2$RuO$_4$-Sr$_3$Ru$_2$O$_7$ eutectic
system, exhibit unconventional SC features. The SC state seems to
be affine with that of Sr$_2$RuO$_4$ because there occurs no
negative interference in the supercurrent flow at the single
interface between Sr$_2$RuO$_4$ and Sr$_3$Ru$_2$O$_7$
macrodomains. The microstructural analysis gives a clear
indication of the amount and size of Sr$_2$RuO$_4$ or other
spurious phases within the Sr$_3$Ru$_2$O$_7$ matrix. Such
information excludes a simple percolation via a proximity network
to justify the observed supercurrent and points the attention on
the role of nanometric RuO stacking faults monolayer in assisting
(driving) the pair leaking (formation).

We would like to acknowledge valuable discussions with C. Noce.

\end{document}